\documentclass[journal=amlccd,layout=twocolumn]{achemso}
\usepackage{epsfig,amsmath,amssymb}
\usepackage{bookmark}
\makeatletter
\renewcommand\@seccntformat[1]{}
\makeatother
\usepackage{graphicx}
\usepackage{dcolumn}
\usepackage{bm}
\usepackage{natbib}
\usepackage{color}
\usepackage{abstract}

\newcommand{\commentMP}[1]%
{\textsf{\textcolor{red}{\{#1\}$^{\mathrm{MP}}$}}}

\author{Sara Jabbari-Farouji $^{1,*}$ and Damien Vandembroucq $^{2}$ }
\affiliation{$^{1}$
Institute of Physics, Johannes Gutenberg-University, Staudingerweg 7-9,
55128 Mainz, Germany
}
 \email{sjabbari@uni-mainz.de }
 
 \affiliation{$^{2}$
Laboratoire PMMH, UMR 7636 CNRS/ESPCI Paris/Universite Pierre et Marie Curie/Universite Paris Diderot
}%

\title[Tensile-compressive asymmetry]
  {Chain level insights into tensile-compressive asymmetry in glassy and semicrystalline polymers}

\begin{document}
{\setlength\paperheight {11in}%
 \setlength\paperwidth {8.5in}}

\twocolumn[\maketitle
\begin{onecolabstract}
 Using molecular dynamics simulations, we provide chain-level insights into the dissimilarities  in  rearrangements of polymers under uniaxial tensile and compressive deformation in glassy and semicrystalline samples of varying chain lengths. The organization of  polymers under tension and compression is distinctively different.  The chains align themselves along the tensile axis  leading to a net global nematic ordering of their bonds and end-to-end vectors whereas under compression, the polymers arrange themselves in a plane perpendicular to the compressive axis   resulting in emergence of an anti-nematic ordering of the bonds and the chain end-to-end vectors. 
 Moreover, the degree of polymers unfolding is greater under tension and  they deform less affinely when compared to chains under compression. The difference between the two responses  strongly depends on the chain length and is the largest at intermediate chain lengths. 
\end{onecolabstract}
]
\begin{tocentry}
 
\begin{center}
      \begin{minipage}[h]{\linewidth}
      \epsfig{file=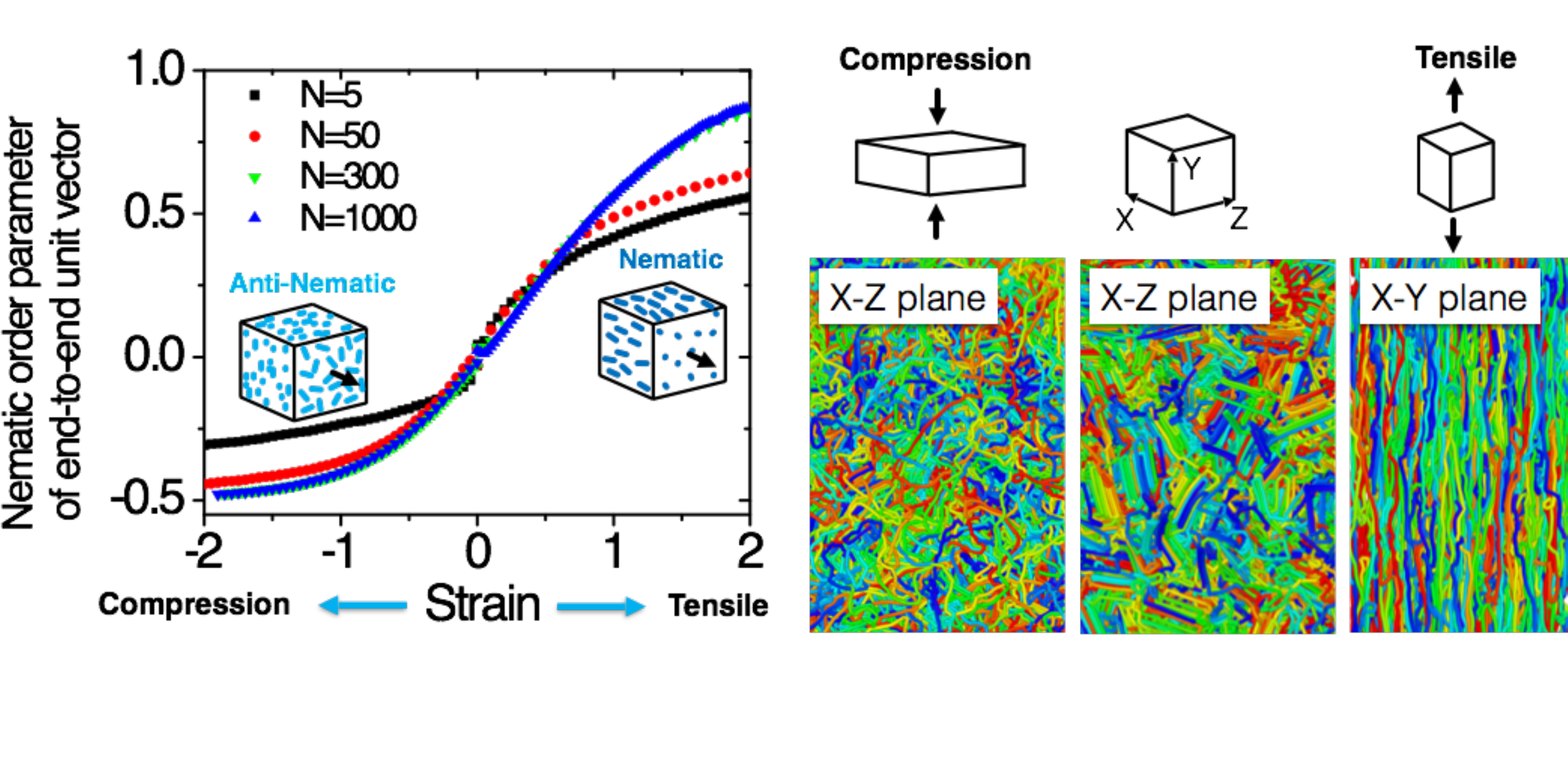, width=0.95\linewidth}
                 \end{minipage}
      \end{center}
Right: The chain end-to-end vectors of semicrystalline polymers develop nematic and anti-nematic ordering under tensile and compressive deformations, respectively.
Left: The conformations of semicrystalline polymers of length 1000, in the undeformed state and under uniaxial tensile and compressive deformations along the $y$-axis.
  
\end{tocentry}





Solid-like polymers used in various high-performance products are
subjected to various loading conditions, {\it e.g.} tensile and
compression. The mechanical response of a polymeric material not only
depends on its structure but also on the imposed deformation mode.~\cite{Mechpolymer1,Mechpolymer2}  For instance, the compressive
yield strengths of polycarbonate and polymethylmethacrylate polymer
glasses are about 20\% larger than their tensile yield strength.~\cite{Asymmetric-glass,PolymerBauschinger}  Therefore, understanding
the link between the deformation-dependent response and  the  underlying
structural rearrangements is crucial for a bottom-up design of
polymeric materials with desired mechanical properties.

 The mechanical responses of glassy and semicrystalline polymers under various deformation modes have been the subject of intensive experimental and theoretical investigations.~\cite{Boyce90,Joerg2003,Hoy2007,MenPRL2003,Humbert2009,Lin2014,Lin2015,PCglassTensilecompression,PolymerBauschinger,glassyplasticflow,Rutledge2011,Rutledge2014,Rutledge2015,Oleinik2018,Micromechcrys,SaraMacro,SaraLMC,SaraPRL2017} However, the effect of deformation mode on conformational and structural arrangement of polymers, especially in the strain-hardening regime, has received much less attention  ~\cite{Boyce90,PCglassTensilecompression,PM-simulation,Rutledge2014}.  Gaining  chain-level insights into deformation mechanisms of polymers is experimentally challenging  due to the small length scales involved. 
Molecular simulations allowing a direct access to various intrachain and interchain statistical measures have played an important role in illuminating the underlying mechanisms  of plastic flow in polymeric solids.~\cite{Hoy2007,Rutledge2011,Rutledge2014,Rutledge2015,polymercomposites,SaraMacro,SaraPRL2017,Joerg2017,Oleinik2018}.Nonetheless, very few of them have focused on the role of deformation mode on these rearrangements.~\cite{PM-simulation,Rutledge2014} Moreover, these studies have used  atomistic simulations that are limited to small samples  due to their high computational costs. For instance, molecular simulations investigating the dependence of deformation mode in semicrystalline polymers focused only on a  layered semicrystalline morphology as part of a larger spherulite structure.~\cite{Rutledge2014} Therefore, the underlying differences in  microstructural arrangements of polymers under various deformation modes and particularly \emph{the effect of chain length} is poorly understood.

To make a headway in this direction, we use large-scale molecular dynamics (MD) simulations of  a crystallizable  bead-spring model ~\cite{Meyer2001} that allows us to generate large samples including up to $2\times 10^6$ monomers. This model is known as  the coarse-grained polyvinyle alcohol (CG-PVA) ~\cite{Meyer2001,melt2018} with a distinctive feature of triple-well bending potential. Upon slow cooling of its melt,  chains undergo  a crystallization transition and form chain-folded  structures  consisting of randomly oriented crystallites with 2D hexagonal order immersed in a network of amorphous strands. For a rapid quench,  polymers retain their amorphous configurations and undergo a glass transition.~\cite{SaraMacro,SaraPRL2017} A prior investigation of tensile
 response of both amorphous and semicrystalline polymers~\cite{SaraPRL2017}  showed  that the response of long polymer glasses
 is dominated by the entanglement network whereas that of
 semicrystalline polymers is determined by an interplay of  the
 two interpenetrated networks of entanglements and tie chains.~\cite{SaraPRL2017}   Here, we focus on the asymmetry between the tensile and compressive  responses of glassy and semicrystalline polymers.

  
 %
\begin{figure*}[t]
\includegraphics[width=0.49\linewidth]{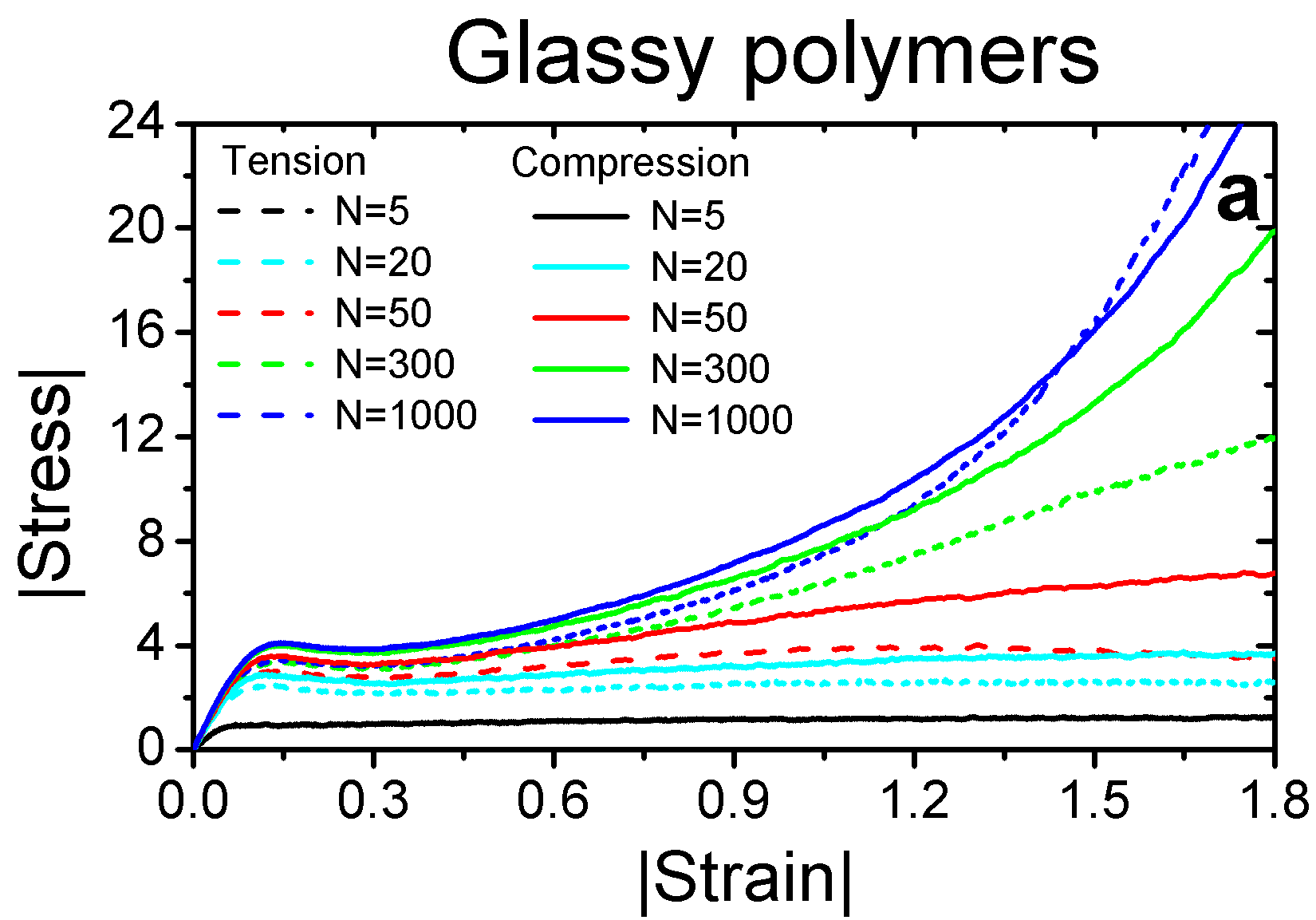}
\includegraphics[width=0.49\linewidth]{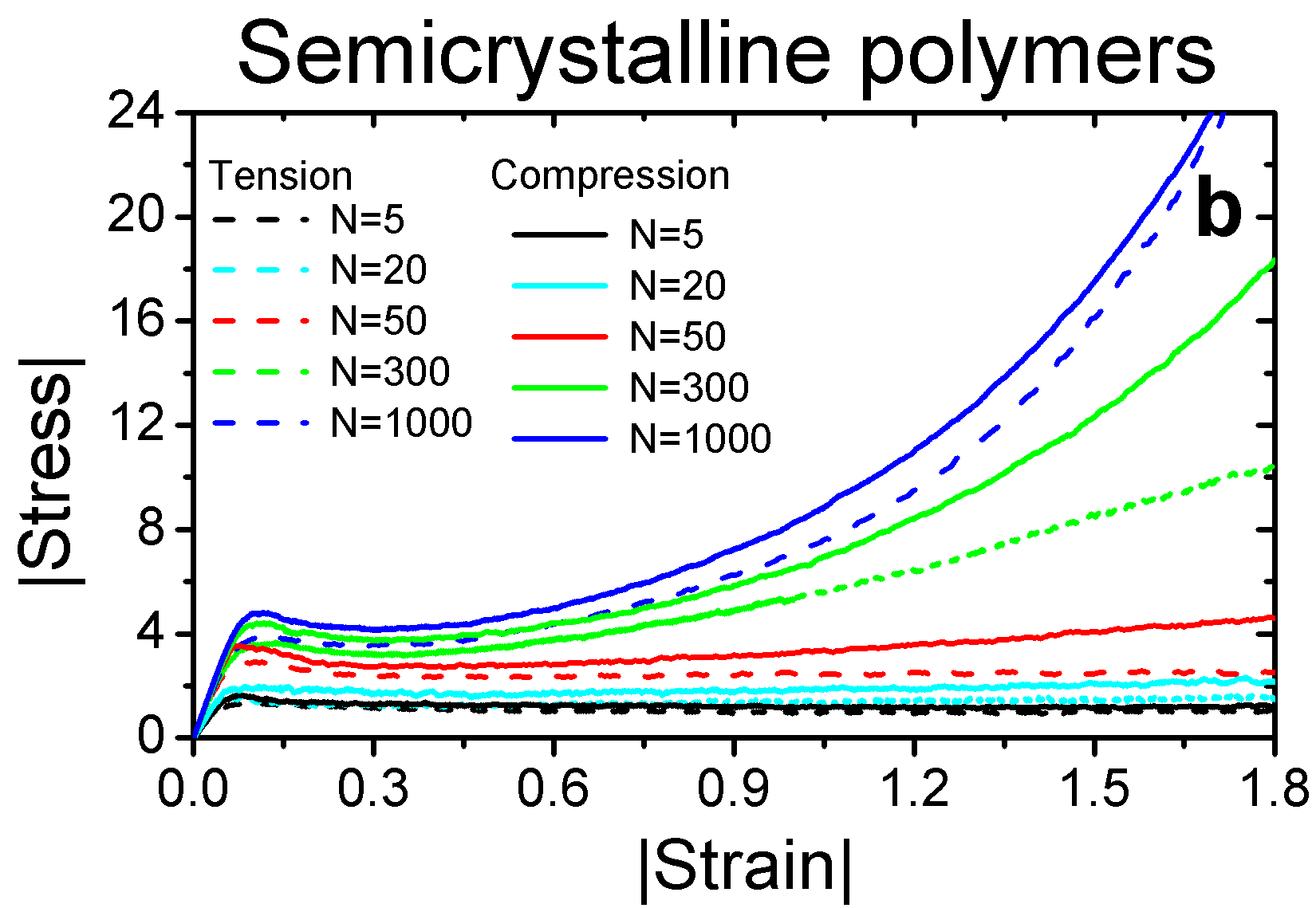}
\caption{
The magnitude of true stress in units of $\epsilon_{\text{LJ}}/ \sigma^3$ versus true strain  from  uniaxial compression (lines)  and  tensile  (circles) deformation of (a)  glassy and (b) semicrystalline   polymers with chain lengths $N=5$, 20, 50, 300  and 1000 as given in the legends.}
 \label{fig1} 
\end{figure*}
 %

  
MD simulations were performed using LAMMPS~\cite{LAMMPS}.  A
  bead-spring model with a triple-well bending potential is
  used.~\cite{Meyer2002,melt2018}
   Distances are reported in units of $\sigma=0.52$~nm, the bond
length is $b_0=0.5 \sigma$. A 6-9 Lennard-Jones potential is used
  to model the non-bonded interactions with the range and strength
$\sigma_{\text{LJ}}=0.89 \sigma$ and $\epsilon_{\text{LJ}}=1.511
k_BT_0$ where $T_0=550 $ K is the reference
temperature~\cite{Meyer2001}. The Lennard-Jones potential is truncated
and shifted at $r_c=1.6 \sigma$. Temperatures $T=T_{real}/T_0$ and
pressures $P=P_{real}\sigma^3/\epsilon$ are reported in reduced
units. 

Glassy and semicrystalline samples of different chain lengths, $N=5$,
20, 50, 300 and 1000,  see ~\cite{SI} for more details, are
obtained from cooling melts at $T=1$ to $T=0.2$ by cooling-rates
$\dot{T}=-10^{-3}\,\tau^{-1}$ and $\dot{T}=-10^{-6}\,\tau^{-1}$,
respectively. The polymer gyration radii span from length scales
comparable to monomers size for $N=5$ to much larger than the
entanglement length in the melt $N_e \approx 36$~\cite{melt2018}.  The
samples are deformed in the $y$-direction with a constant true
strain-rate of $\dot{\varepsilon}=\pm 10^{-5} \tau^{-1}$ while
imposing a constant pressure $P=8$ (as in the undeformed samples) in
the $x$ and $z$-directions.
The volume increase is at most 10 \% for both glassy and
semicrystalline polymers at the largest strain $|\varepsilon|=1.8$ and
$T = 0.2$ ~\cite{SI} and these systems behave nearly
as an incompressible fluid for $|\varepsilon|< 1$.

\begin{figure*}[t]
\includegraphics[width=0.48\linewidth]{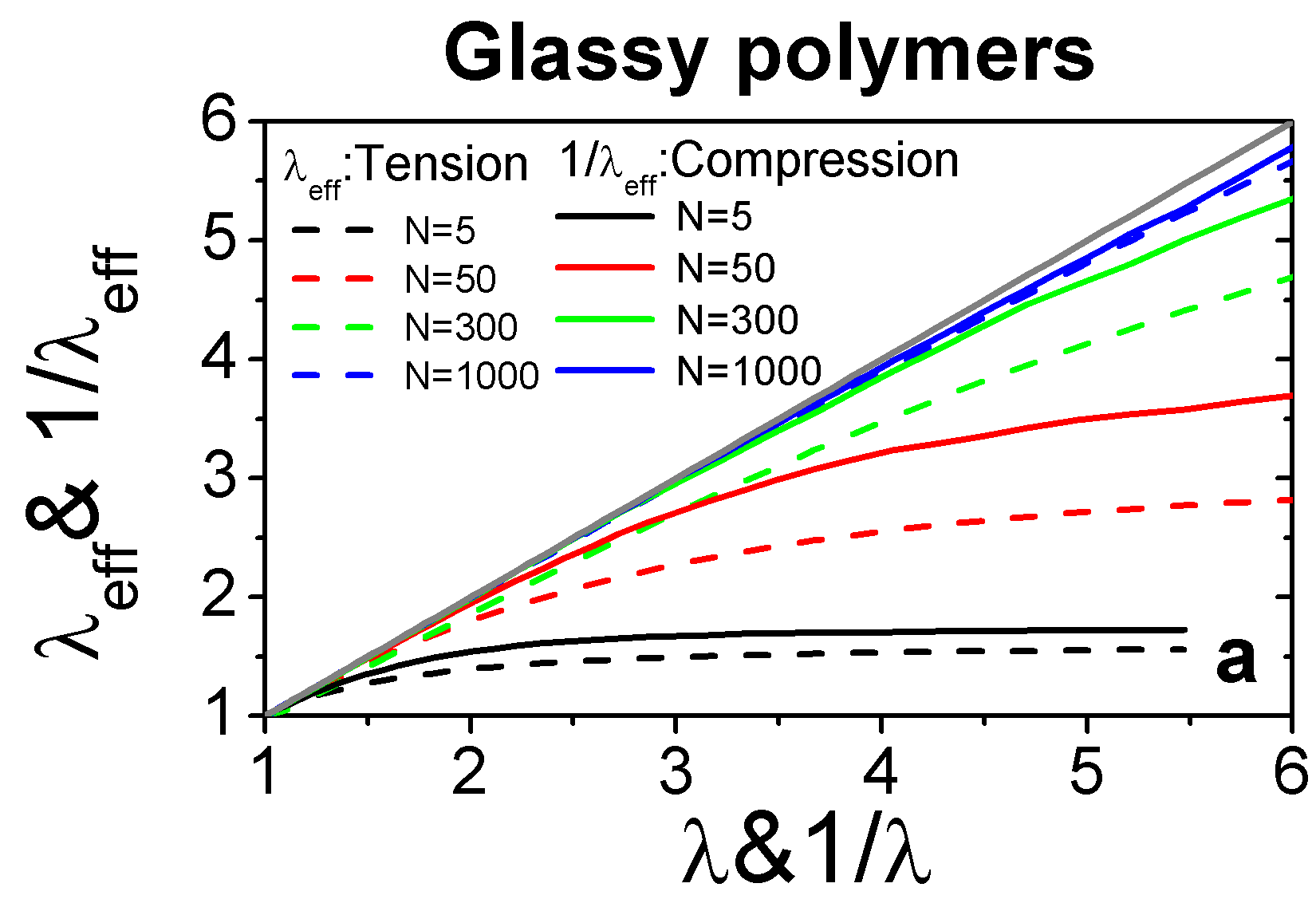}
\includegraphics[width=0.48\linewidth]{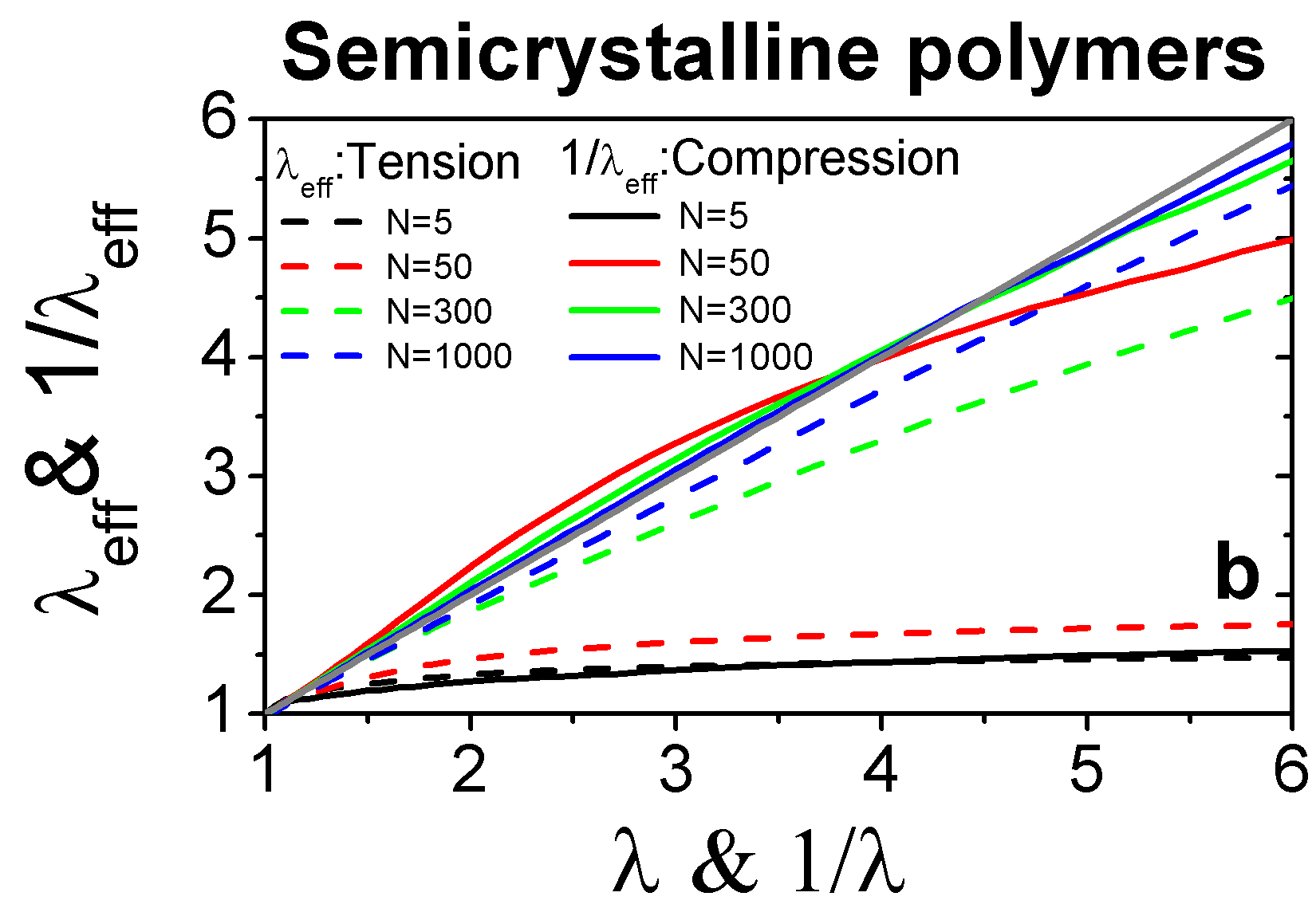}
\caption{Effective microscopic stretch  $\lambda_{eff} \equiv \sqrt{\frac{\langle {R_y}^2 \rangle} {\langle {R_y^0}^2 \rangle}} $ versus macroscopic stretch $\lambda\equiv L_y/L_y^0$  for samples  under  tension (dashed lines) and  $1/\lambda_{eff}$ versus $1/\lambda$ for polymers under compression (solid lines)  in (a) glassy and  (b) semicrystalline polymers with different chain lengths as given in the legends. }
\label{fig2} 
\end{figure*}
Figures \ref{fig1}a and b present stress-strain curves for glassy and
semicrystalline polymers obtained under compressive and tensile
deformation, respectively.  For all the samples, the elastic regime at
small strains is followed by a plastic flow at larger deformations.
For the shortest chain length $N=5$ with $R_g \approx 0.7\sigma$, we
observe a stress plateau beyond the yield point and the compressive
and tensile responses of both glassy and semicrystalline polymers are
almost identical. For longer polymers, the stress response is larger
in compression than in tension. The asymmetry is most pronounced at
intermediate chain lengths $N=50$ and 300. Interestingly, for the
longest chain length $N=1000$, the compressive and tensile responses
become similar again.

To gain insights into chain rearrangements under macroscopic
deformation, we measure the RMS components $R_i$ of the end-to-end
vectors of chains as relative to their initial values $R^{\alpha}_0$.~\cite{Hoy2007} The effective microscopic stretch ratio is defined
as $\lambda_{eff} \equiv \sqrt{\frac{\langle {R^y}^2 \rangle} {\langle
    {R^y_0}^2 \rangle}}$ where the $y$-direction is taken as the
tensile/compressive axis.  Under an affine uniaxial compressive or
tensile deformation, $\lambda_{eff}$ is identical to the macroscopic
stretch ratio obtained as $\lambda\equiv L_y/L_y^0$ and the changes in
$\langle{R^{\alpha}}^2 \rangle$ are consistent with a volume
conserving uniaxial macroscopic deformation, {\it i.e.},
$\lambda_{eff}=\frac{\langle {R^x_0}^2 \rangle} {\langle {R^x}^2
  \rangle}=\frac{\langle {R^z_0}^2 \rangle} {\langle {R^z}^2
  \rangle}$.  The ratio $\lambda_{eff}$ has recently been
  identified as  an important parameter controlling the strain
hardening behavior of amorphous and semicrystalline
polymers~\cite{Hoy2007,SaraPRL2017}  in the sense
  that, independently of the chain lengths, when $\lambda_{eff} (\lambda)$ of two
  samples are similar, their responses in the strain-hardening regime
  are also alike.
 
 Figures \ref{fig2}a and b display $\lambda_{eff}$ as a function of
 the macroscopic stretch $\lambda$ for tensile measurements and
 $1/\lambda_{eff}$ versus $1/\lambda$ for compressive tests of glassy
 and semicrystalline polymers, respectively. Similar to the results
 for tensile deformation, ~\cite{SaraPRL2017} under compression
 polymers in semicrystalline state deform less affinely than their
 glassy counterparts. The general trend that we observe is that the
 longer chains exhibit a more affine behavior irrespective of
 deformation mode and their underlying structure. Overall, polymers
 deform more affinely under compression than under tension.
 Notably, the effective microscopic stretch reflects well the
 compressive-tensile asymmetry observed in the stress-strain curves in
 Fig. \ref{fig1}. The difference between $\lambda_{eff}$ in the tensile
 deformation and $1/\lambda_{eff}$ in the compression is  large  when
 the asymmetry between the two responses is significant.

\begin{figure}[h]
\includegraphics[width=0.49\linewidth]{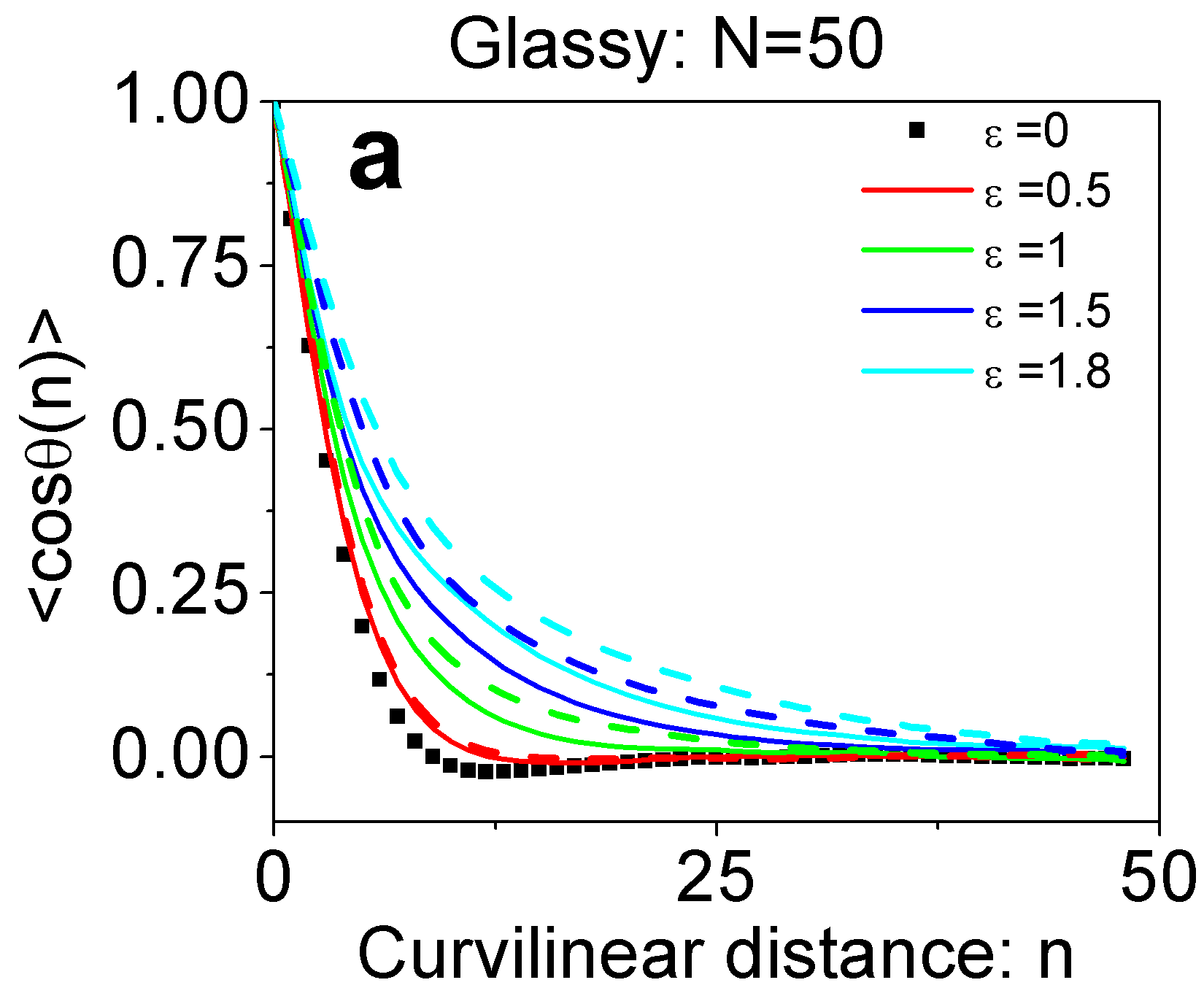}
\includegraphics[width=0.49\linewidth]{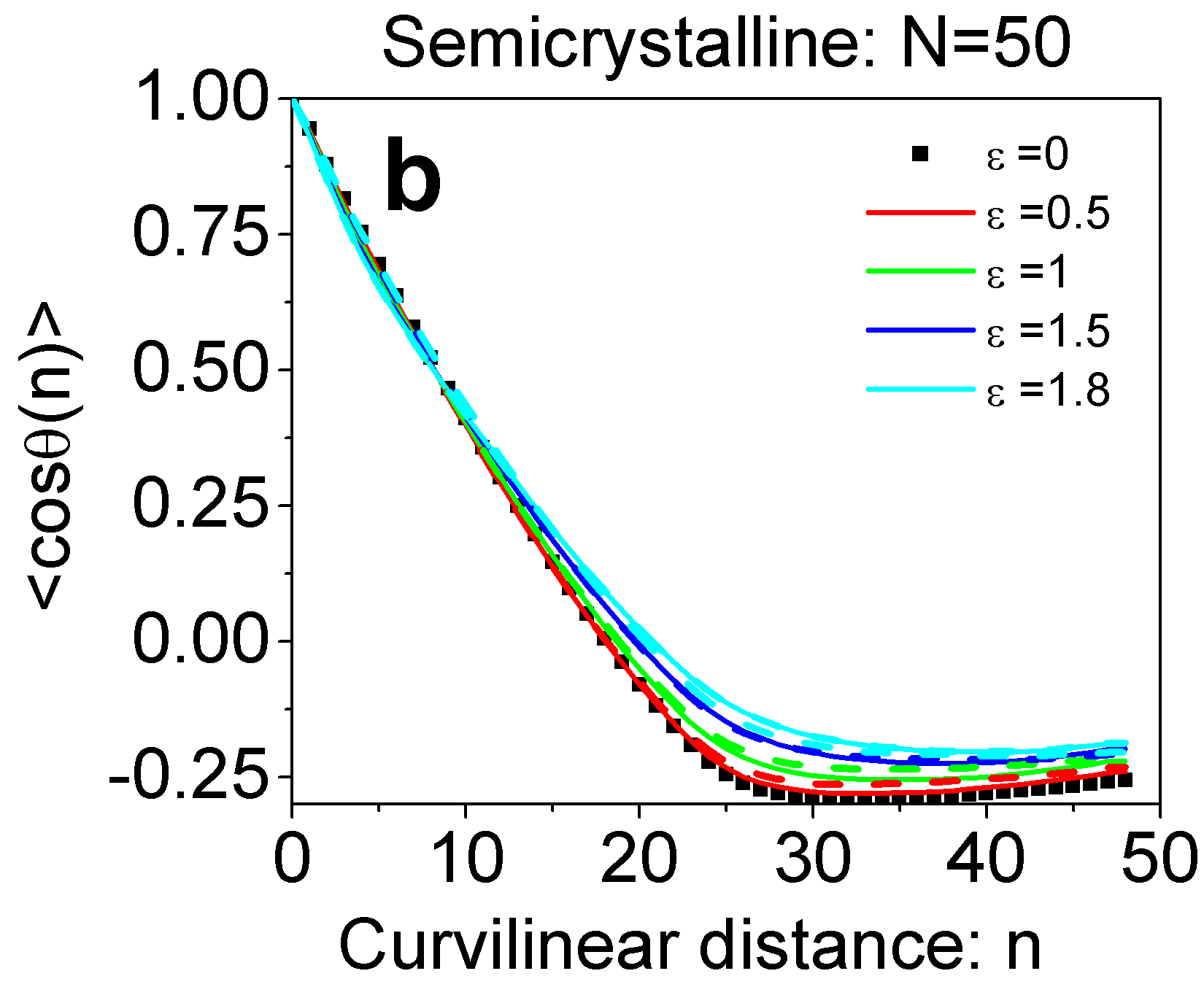}
\includegraphics[width=0.49\linewidth]{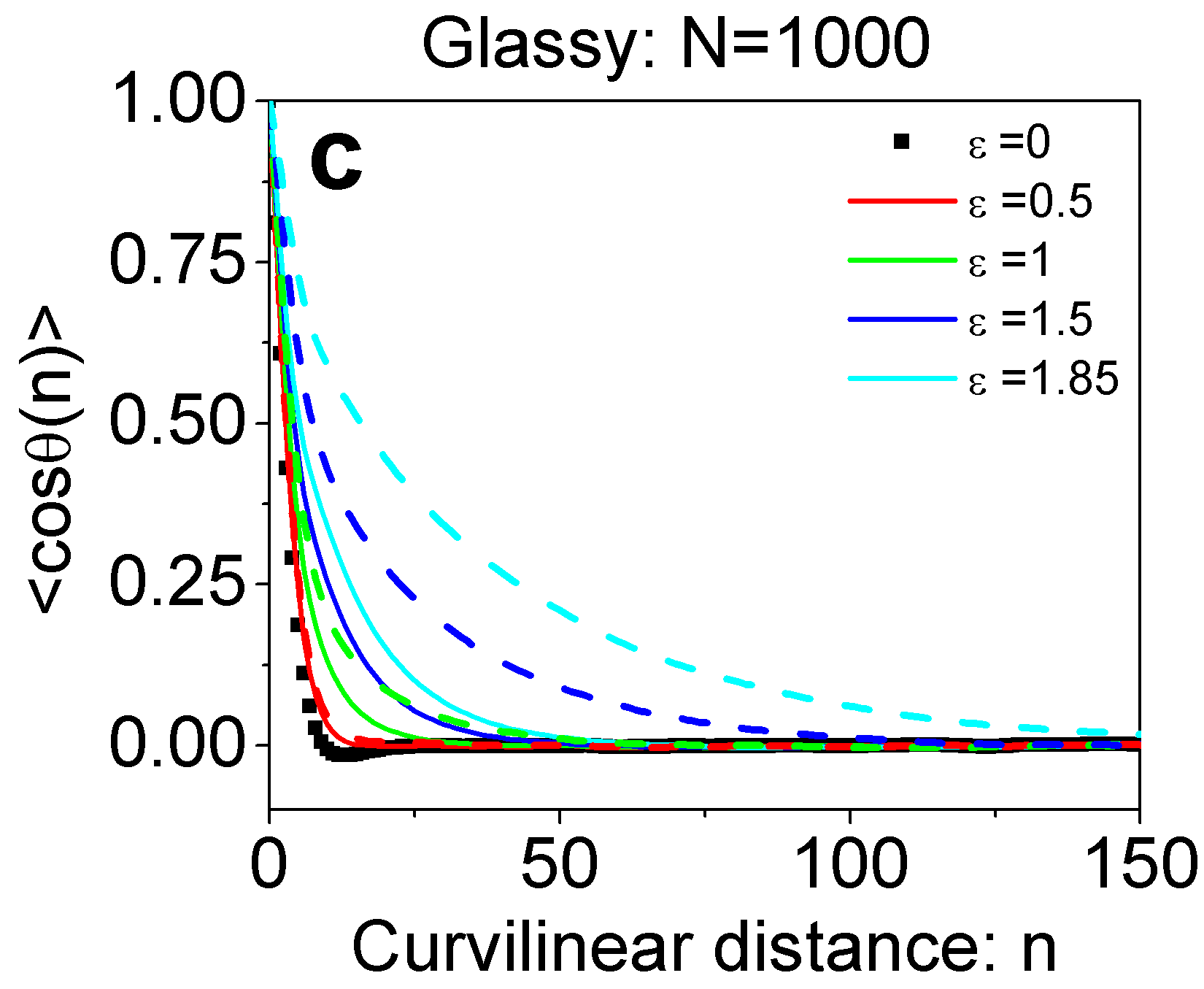}
\includegraphics[width=0.49\linewidth]{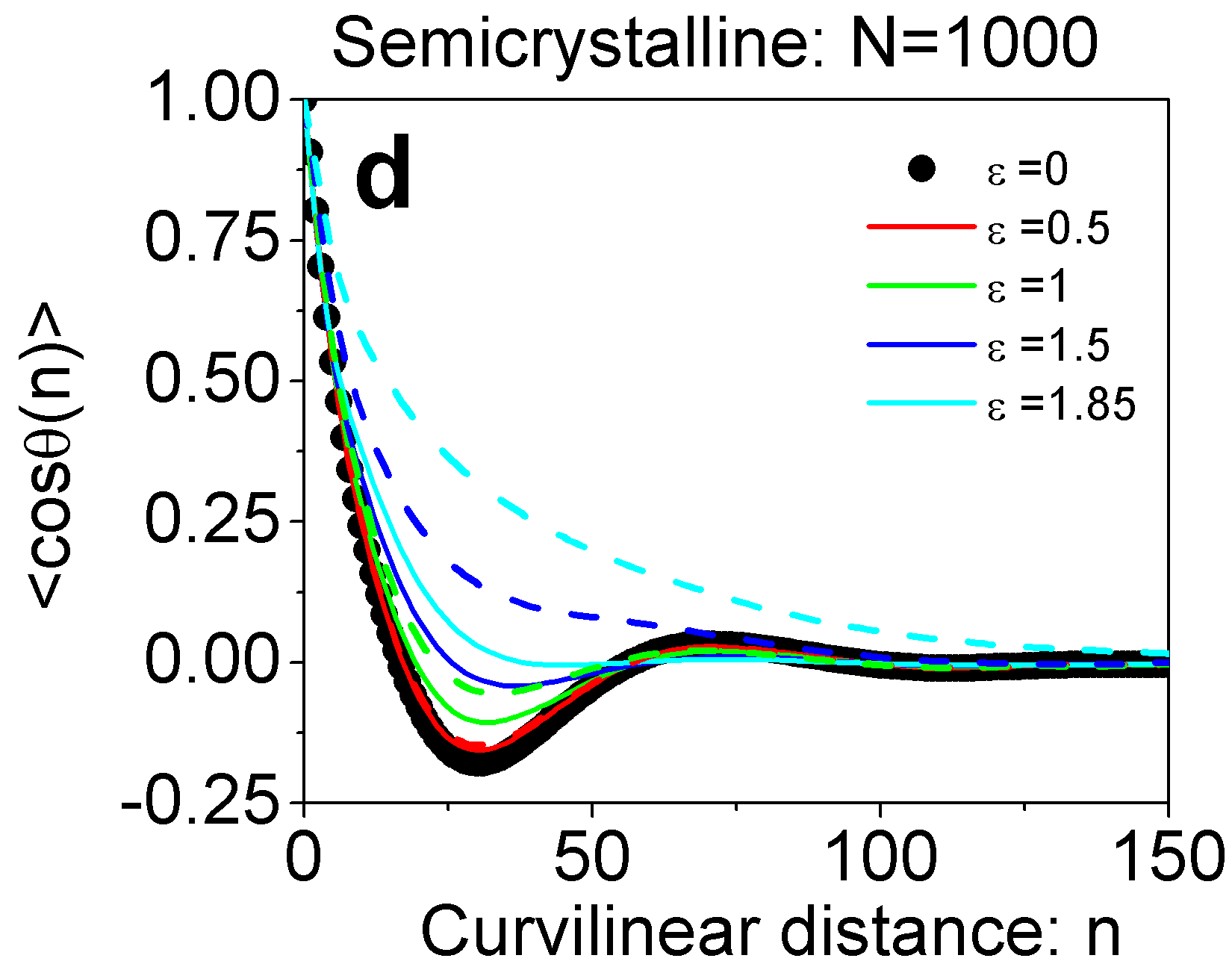}
\caption{Evolution of intrachain bond-bond orientational correlation function $ \langle \cos \theta  (n) \rangle$  against the curvilinear distance $n$ plotted  at different stages of compressive (solid lines) and tensile (dashed lines) deformation for glassy and semicrystalline polymers with chain  lengths $N=50$ and 1000. The  strain magnitudes $\varepsilon$ are given in the legends.}
\label{fig3} 
\end{figure}

 To directly quantify the evolution of the average chain conformations under
 deformation, we compute the intrachain orientational bond-bond
 correlation function. It is defined as $ \langle \cos \theta (n)
 \rangle \equiv \langle \widehat{\mathbf{b}}_{i,j} \cdot
 \widehat{\mathbf{b}}_{i,j+n} \rangle $ where
 $\widehat{\mathbf{b}}_{i,j}$ denotes the orientation vector of the
 $j$th bond in the $i$th chain and $1 \le n \le N-1$ is the
 curvilinear distance between any two monomers in a
 chain. Fig. \ref{fig3} represents $ \langle \cos \theta (n) \rangle$
 at different stages of compressive and tensile deformations for
 glassy and semicrystalline samples with chain lengths $N=50$ and
 $N=1000$. Upon increase of strain and  chain unfolding, the
 intrachain orientational correlation length grows  regardless of
   the initial structure and deformation mode. Nonetheless, under
 identical conditions the degree of chain unfolding in the tensile
 deformation is larger compared to the compression.

\begin{figure}[t]
\includegraphics[width=0.49\linewidth]{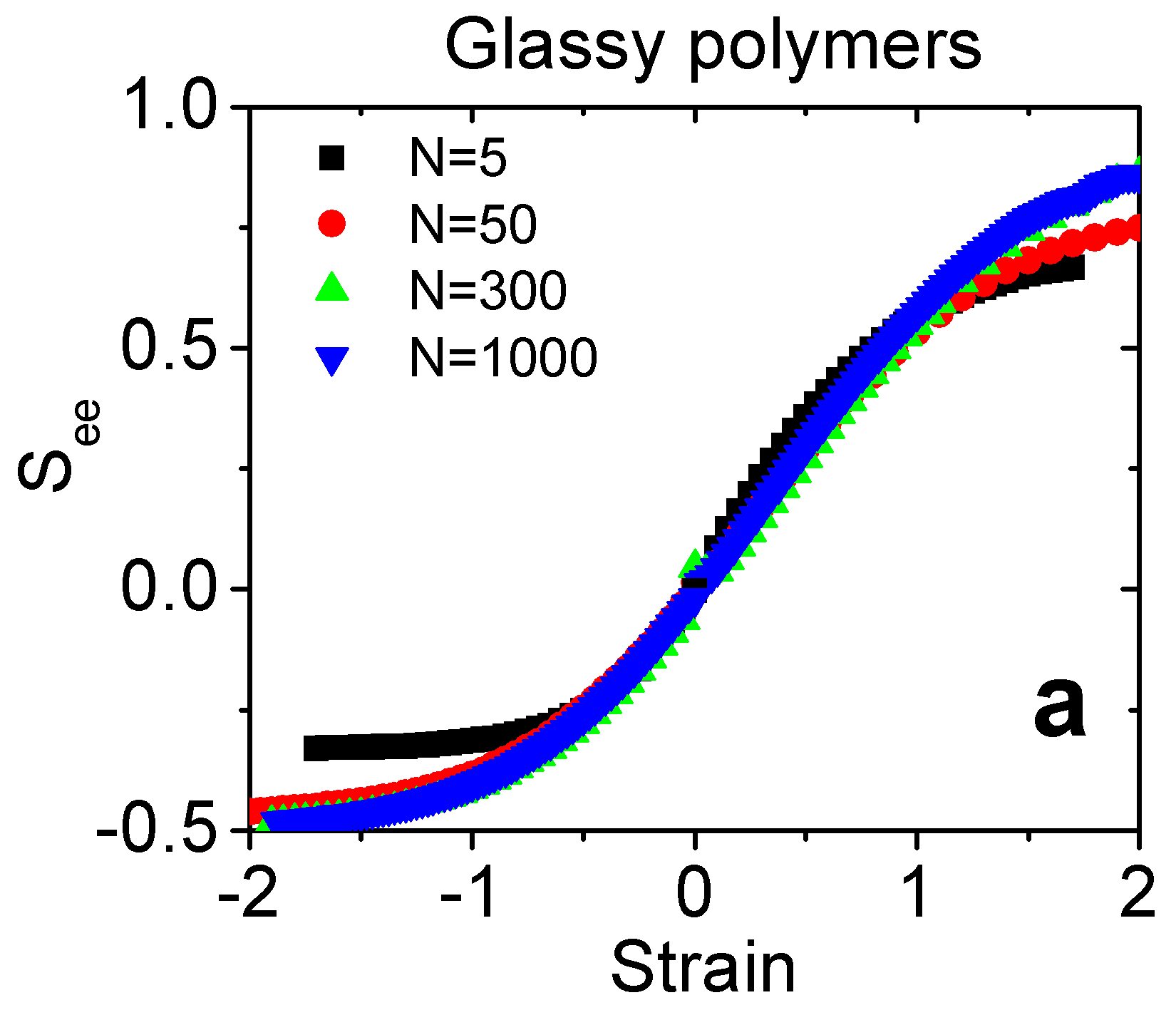}
\includegraphics[width=0.49\linewidth]{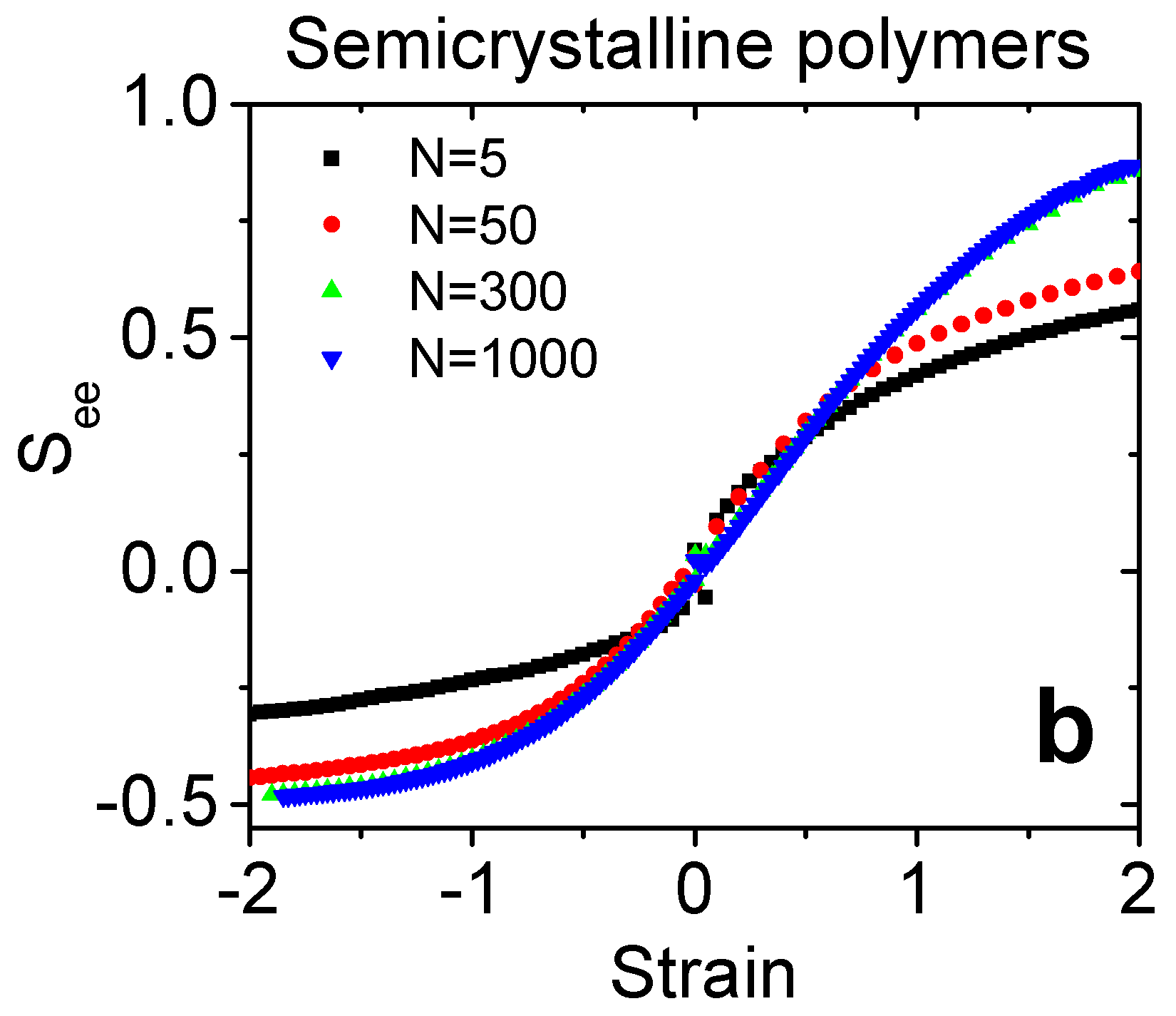}
\includegraphics[width=0.49\linewidth]{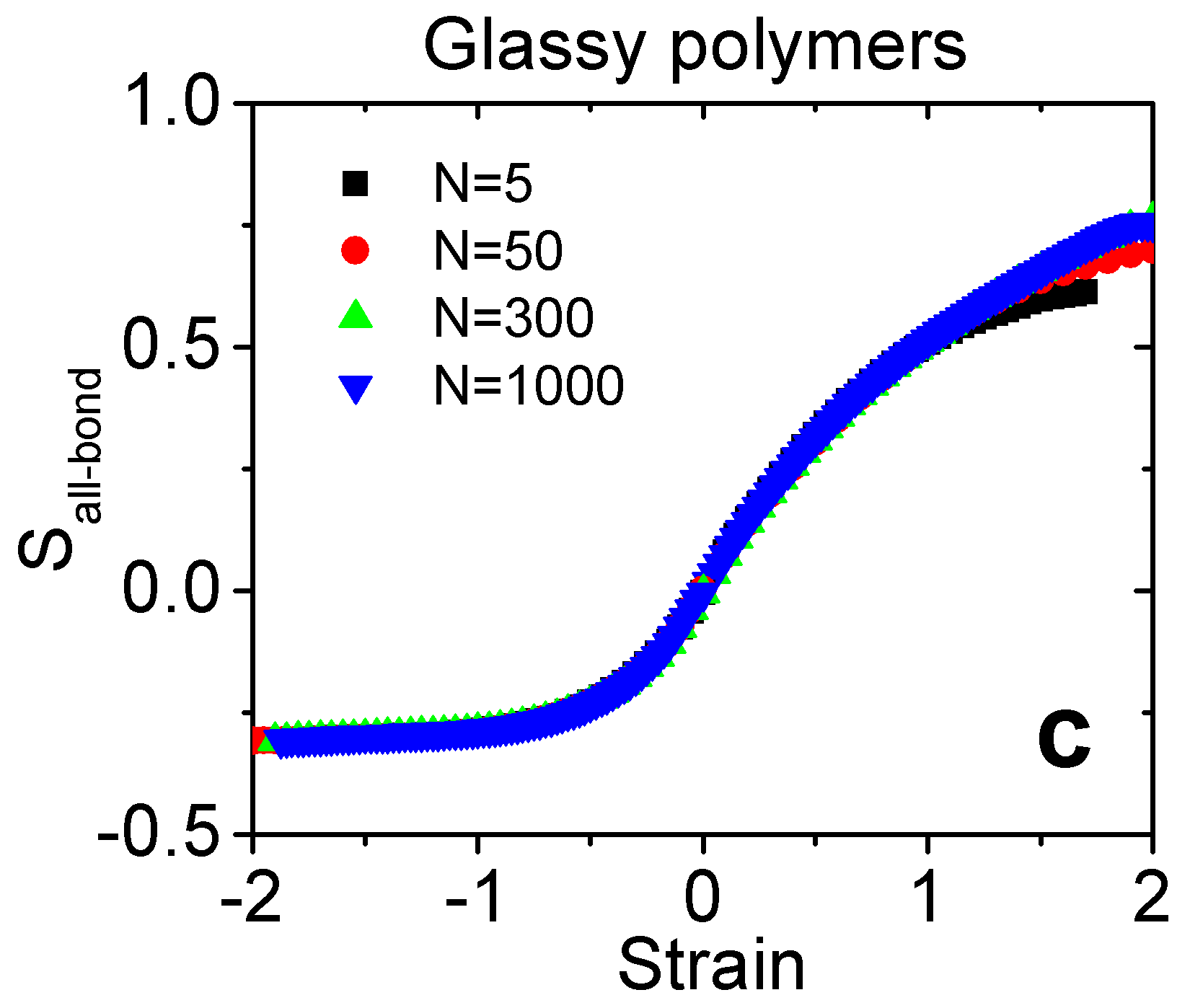}
\includegraphics[width=0.49\linewidth]{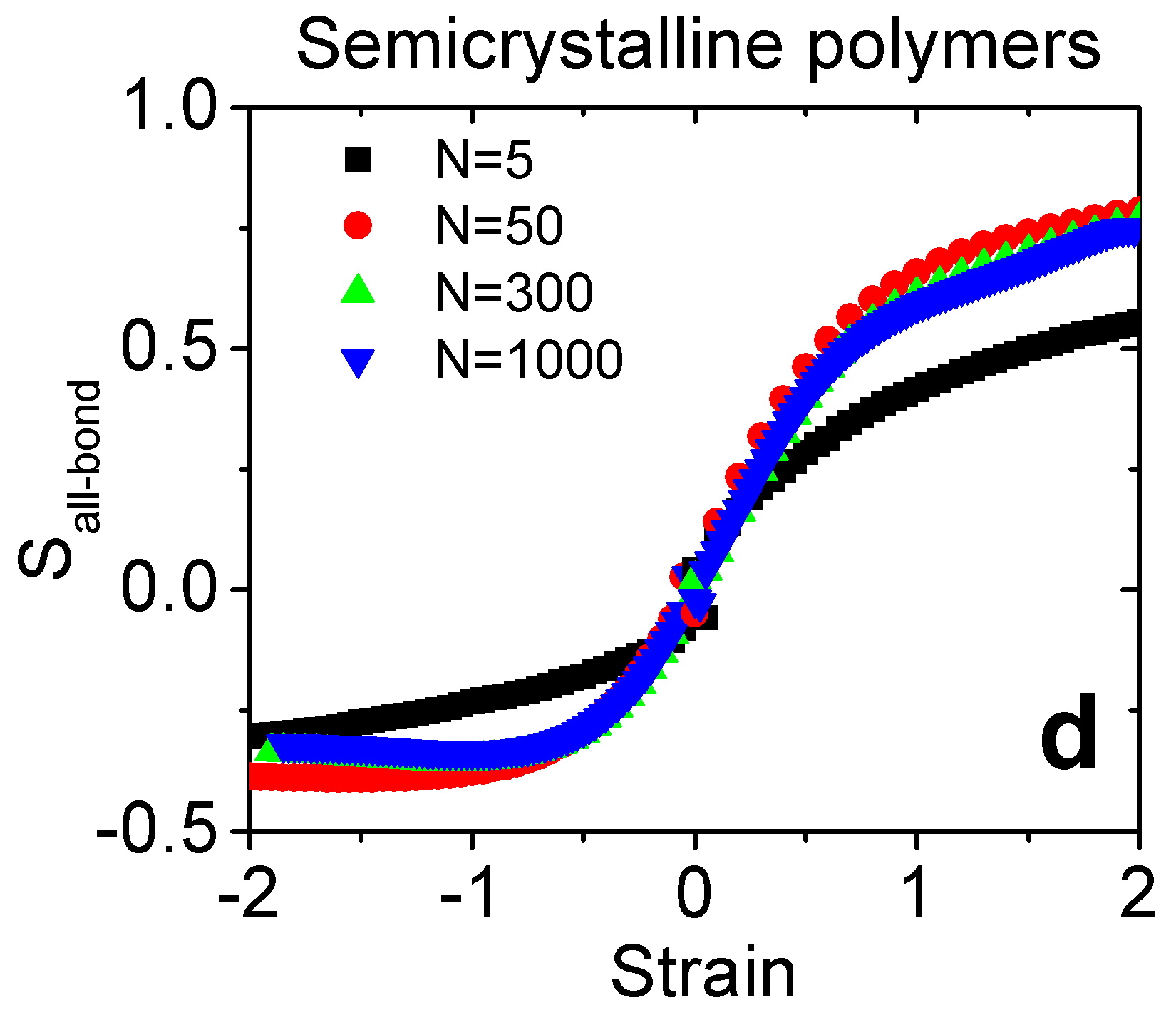}
\caption{The nematic order  parameter of chains end-to-end vector, $S_{ee}$ for glassy (a) and semicrystalline (b)  polymers of different chain lengths, $N$, as a function of strain in both compression (negative strain) and tensile deformation modes.  Likewise, the strain-dependence of   the global nematic order parameter of all bonds, $S_{\text{all-bond}}$, for glassy (c) and semicrystalline (d) polymers. }
\label{fig4} 
\end{figure}
%
We next examine
the collective organization of the chains.  To this end, we calculate
the nematic order tensor of all the chain end-to-end orientation
vectors $\widehat{\mathbf{R}}_i$ obtained as $\mathbf{Q}=\frac{1}{2
  n_{c} }\sum_{i=1}^{n_{c}} (3 \widehat{\mathbf{R}}_i
\widehat{\mathbf{R}}_i- \mathbf{I})$ where $n_{c}$ is the the number
of chains in a sample. We define $S_{ee}$ as the eigenvalue of
$\mathbf{Q}$ with the largest  \emph{magnitude} varying in the range $ -0.5\le
S_{ee} \le 1$.  In the isotropic case, the three eigenvalues are
  null and $S_{ee}=0$. The degeneracy is lifted as soon as the
  structure gets anisotropic. Two extreme cases can be considered.
When all the end-to-end vectors are aligned in the same direction,
$S_{ee} \rightarrow 1$ leading to a perfect uniaxial nematic alignment. On the
other hand, when all the end-to-end vectors are perpendicular to a
direction, a uniaxial anti-nematic order charactrized by $S_{ee} \rightarrow
-0.5$ emerges.  Fig.~\ref{fig4}a and b show the evolution of $S_{ee}$
as a function of strain for glassy and semicrystalline polymers,
respectively.  Under tensile deformation, the chains end-to-end
vectors get aligned along the tensile axis and $S_{ee}$ progressively
increases with strain, approaching unity at large strains. On the
contrary under compression where the strain is negative, $S_{ee}$
becomes gradually more negative and it approaches $-0.5$ already at
$\varepsilon \approx -1.5$ for $N>5$. This anti-nematic ordering reflects
a distinct arrangement of chains under compression.  As they are compressed
 in one direction, the chains elongate isotropically in directions perpendicular to the compressive axis,
and their end-to-end vectors lie in a plane normal to the deformation axis.   Likewise, we define $S_{\text{all-bond}}$ as the largest magnitude eigenvalue of
the global nematic tensor of all the bond orientation vectors $\mathbf{Q}=\frac{1}{2
 n_{\text{bond}} }\sum_{i=1}^{n_{\text{bond}}} (3 \widehat{\mathbf{b}}_i
\widehat{\mathbf{b}}_i- \mathbf{I})$ where $n_{\text{bond}}$ is the the number
of all the bonds in a sample.
Investigating the global nematic order parameter of the bond
vectors $S_{\text{all-bond}}$ presented in Fig.~\ref{fig4}c and d, we
observe a very similar trend; nematic ordering under tensile
deformation and anti-nematic ordering under compressive deformation.

  Our results highlight the distinct nature of inter-  and intra-chain organization in glassy and semicrystalline polymers under compressive and tensile deformations. In both cases, the initial
hexagonal order of semicrystalline polymers is destroyed at large deformations and the configurations of semicrystalline and glassy polymers become similar. Under tension, the chain end-to-end vectors and the bond vectors of all the samples
   align themselves along the deformation axis leading to a net nematic ordering  whereas under compression a
  novel anti-nematic ordering emerges as a result of chain unfolding.  Such an anti-nematic
  ordering suggests that a buckling-like mechanism may be at play under compression as supported by visual inspections and distribution of bond angles.~\cite{SI}
  The two responses become similar for very short and long chains, whereas the asymmetry is
  maximum  at intermediate chain lengths. As mentioned earlier, short chains behave like monomers.  Upon increase of chain length beyond the entanglement length $N_e\approx 36$, an entanglement  network builds up in glassy polymers. In semicrystalline polymers,  part of the entanglement network is lost but instead a network of tie chains connecting the crystalline domains is developed  where  the average tie-chain length varies in the range $10<L_{\text{tie}} <30$ for $50 \le N <1000$ \cite{SaraPRL2017}. Hence, the longest chains $N=1000$ generate long enough tie chains comparable to their  entanglement length $N_e\approx 40$ ~\cite{Luo2013} that enforce affine
deformation at the chain level whereas intermediate chain lengths deform non-affinely. 
The effect of deformation mode on the entanglement network structure and the distribution of entanglement
  points along the chains remains an open question that merits future investigations. 

 \begin{acknowledgement}
 S.~J.-F. acknowledges financial support from the German Research Foundation (http://www.dfg.de) within SFB TRR 146 (http://trr146.de). 
The computations were performed using the Froggy platform of the CIMENT infrastructure supported by the Rhone-Alpes region (Grant No. CPER07-13 CIRA) and the Equip@Meso Project (Reference 337 No. ANR-10-EQPX-29-01) and  the supercomputer clusters Mogon I and II at Johannes Gutenberg University Mainz (hpc.uni-mainz.de).
 \end{acknowledgement}
 
 \bibliography{polymer2.bib}

\end{document}